\documentclass[%
  aps,
  pra,
  reprint,
  superscriptaddress,  
  longbibliography,    
  nofootinbib
]{revtex4-2}

\usepackage{graphicx}  
\usepackage{amsmath,amsfonts,amssymb}  
\usepackage{siunitx}   
\newcommand{\affilLL}[0]{Lincoln Laboratory, Massachusetts Institute of Technology, Lexington, Massachusetts 02421, USA}
\newcommand{\affilMIT}{Massachusetts Institute of Technology, Cambridge, Massachusetts 02139, USA}

\usepackage{hyperref}  
\usepackage[nameinlink, capitalize]{cleveref}  
\usepackage{braket}    

\begin{document}

\title{Real-time magnetic field noise correction using trapped-ion monitor qubits}

\author{Kyle DeBry}
\affiliation{\affilMIT}
\affiliation{\affilLL}
\email{debry@mit.edu}

\author{Agustin Valdes-Martinez}
\affiliation{\affilMIT}
\affiliation{\affilLL}

 \author{David Reens}
\affiliation{\affilLL}

\author{Colin D. Bruzewicz}
\affiliation{\affilLL}



\author{John Chiaverini}
\affiliation{\affilMIT}
\affiliation{\affilLL}

\date{\today}

\begin{abstract}
We demonstrate a trapped-ion protocol in which a nearby, dedicated ``monitor'' qubit tracks magnetic-field drifts in real time without interrupting data-qubit operations. Using two $^{40}\mathrm{Ca}^+$ ions and the optical--metastable--ground architecture, we encode the data qubit in the ground-state manifold and the monitor qubit in a metastable-state manifold to achieve spectral separation. The monitor qubit senses common magnetic fluctuations during data-qubit experiments, enabling feedforward corrections to the qubit-control drives. Under applied magnetic noise with a realistic spectrum ($1/f^{2}$), the protocol maintains coherence and, when compared with interleaved calibration, it extends usable data-qubit probe times by up to a factor of ${\sim}\sqrt{2}$ and doubles the experimental duty cycle. These results establish monitor qubits as a scalable tool for real-time recalibration in quantum information processors.
\end{abstract}

\maketitle

\section{Introduction}
\begin{figure*}
    \centering

    \includegraphics[width=\linewidth]{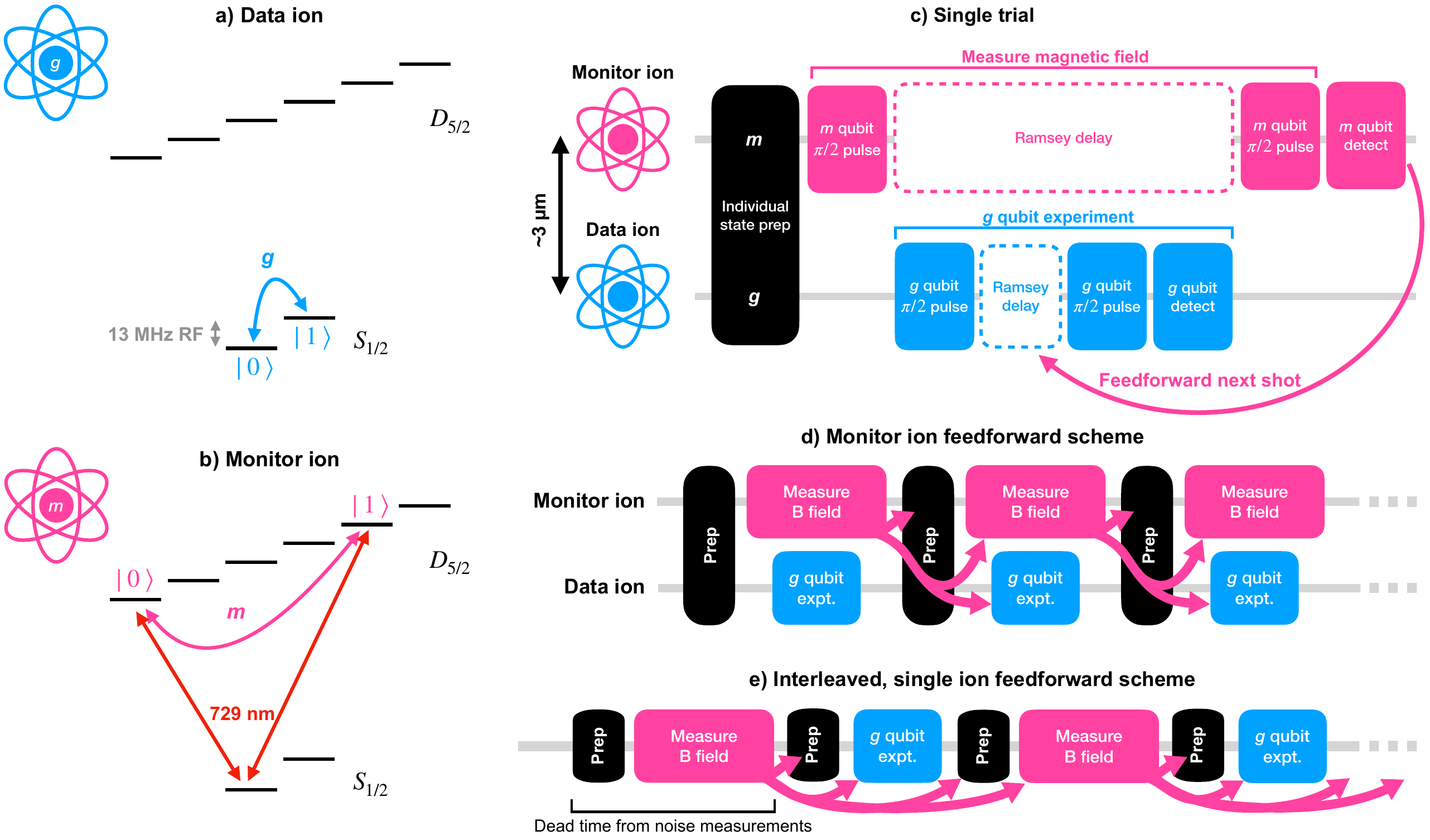}

    \caption{Monitor ion experiment overview. \textbf{a)} Energy levels used for the data ion, encoded as a $g$-type qubit and addressed with \SI{13}{\mega\hertz} RF radiation. \textbf{b)} Energy levels used for the monitor ion, encoded as an $m$-type qubit and addressed with \SI{729}{\nano\meter} laser light. \textbf{c)} Circuit diagram of one realization of the experiment, including the state preparation pulse, the monitor ion Ramsey sequence to measure the magnetic field, and the data ion's experiment (implemented here as a simple Ramsey sequence). Pulse durations are not drawn to scale. \textbf{d)} Circuit diagram showcasing the feedforward scheme, where the information obtained from the monitor ion's magnetic field measurement is used to inform the pulse sequence applied during the following data ion experiment realization. Prep: state preparation. \textbf{e)} Circuit diagram of the interleaved case used for comparison, where a single ion alternates between experiments and magnetic-field calibration measurements. Additional dead time is accrued in this case compared to the separate monitor qubit protocol due to the noise measurements occurring in series with the data qubit experiments, rather than in parallel (times not drawn to scale).}
    \label{fig:monitor-ion-qubits}
\end{figure*}
Mitigating the impacts of external noise sources is a primary challenge to building a useful quantum computer \cite{wineland1998experimental, Nielsen2010}. While quantum error correction will likely be required, it can only help in situations in which the noise is sufficiently small that the fidelity of each operation is above the error correcting code's threshold. While there are numerous methods for passively mitigating noise in quantum systems~\cite{langer2005long, Rotheudt2025, Kornack2007, Liu2024, Chupp2019}, it is likely that active noise measurement and frequent recalibration will also be needed \cite{Wei2022, Belfi2010, Tiengo2025, Merkel2019}.


To perform long, fault-tolerant algorithms, quantum computers will require advanced methods to measure and correct for drifts in control parameters. If drifts are not continuously monitored, they will lead to coherent errors that accumulate and degrade the fidelity of all aspects of quantum computer operation. Traditional recalibration techniques often rely on periodic, offline characterization routines that take up valuable computational time. Feedback from classical sensors can also be used, but is often insufficiently correlated with the qubits' local environment to ensure high-fidelity operation. In processing campaigns that may take days or longer to complete a full fault-tolerant algorithm \cite{Gidney2025}, the quantum circuit cannot be halted in order to run recalibration routines on the same qubits that are being used to store quantum data. Without a mechanism for real-time, local monitoring of drifts, quantum processors are susceptible to increased error rates, compromising their performance.


Dedicated monitor qubits (also sometimes referred to as ``spectator'' qubits~\footnote{We use \textit{monitor qubit} to avoid the ambiguity posed by \textit{spectator qubit}, as the latter term can also be used to refer to a qubit that, while not involved in a particular computation, is nevertheless coupled to the involved qubits or control fields, generally leading to deleterious consequences \cite{tripathi2022suppression}.}) provide a scalable and robust solution to these recalibration challenges by enabling continuous, in-situ monitoring of experimental drifts \cite{Singh2023, Majumder2020, Gupta2020b, Tonekaboni2023, Song2023, Loenen2025, Liu2025, Danageozian2023, Zheng2024, Danageozian2022}. By reserving a subset of qubits solely for sensing critical parameters, monitor qubit protocols enable continuous recalibration without disrupting the algorithm performed on the data qubits. Real-time analysis of monitor-qubit readouts enables immediate feedforward corrections to the pulses applied to data qubits, reducing errors due to environmental fluctuations. 


While theoretical aspects of monitor qubits have been the subject of a number of studies, there have only been a few experimental implementations. Experiments in arrays of neutral atoms achieved spectral selectivity by using atoms of different species for data and monitor qubits, showing the ability to counteract monochromatic magnetic field noise~\cite{Singh2023}. Atomic clock experiments have similarly used multiple ensembles to address laser phase noise \cite{Clements2020, Kim2023, Zheng2024}. However, these demonstration have addressed only global effects, and do not address localized noise. Recently, an experiment using nitrogen-vacancy centers in diamond used multiple nuclear spins coupled to a single electron, but the approach is limited by crosstalk when accessing the different qubits \cite{Loenen2025}.

Multi-species control methodologies have also been used fruitfully in trapped-ion systems, for applications such as sympathetic motional cooling, but come at the cost of increased experimental complexity. Alternatively, one can make use of the large Hilbert space available in atomic systems to achieve similar spectral selectivity with only a single atomic species \cite{Allcock2021}. 
The use of spectrally-separated monitor qubits, such as dual-species arrays \cite{Singh2023, Zheng2024} or through \textit{omg} (optical-metastable-ground), or dual-type, architectures \cite{Allcock2021, Yang2022, Feng2024, Lis2023, Chen2022}, ensure that the act of sensing does not induce crosstalk, making continuous recalibration compatible with extended quantum computations.

Here we use the \textit{omg} paradigm to implement a single-species monitor qubit scheme with two trapped ions to mitigate detuning errors that would otherwise be induced by magnetic-field drifts.  Applying broadband magnetic-field noise of varying amplitude, we correct the phase of operations on a data ion on every realization based on measurements of a co-trapped monitor ion, maintaining coherence through real-time tracking of the noise-induced detuning errors.  We also demonstrate an increase in both the usable data-qubit probe time and the effective duty cycle of experiments with recalibration, in light of realistic environmental perturbations.  These methods, adapted from single-ion atomic-clock interrogation techniques~\cite{bergquist1991single, Fisk1997, bernard1998laser, madej2000single, Ludlow2015}, are widely applicable across various noise sources and qubit types.

\section{Results}
Two calcium ions are trapped in a single harmonic well in a cryogenic surface-electrode trap, described previously \cite{Sage2012, Bruzewicz2016, DeBry2023, DeBry2025}. Prior to each realization of the experiment, the ions are cooled to the Doppler limit with \SI{397}{\nano\meter} light and subsequently optically pumped to the $m=-1/2$ sublevel of the $S_{1/2}$ ground state. A \SI{729}{\nano\meter} laser used to drive electric-quadrupole $S_{1/2} \leftrightarrow D_{5/2}$ transitions is stabilized to an ultra narrow-linewidth reference cavity. An in-vacuum pair of wire coils oriented normal to the quantization axis delivers RF fields to the ions to drive magnetic-dipole transitions between the $S_{1/2}$ Zeeman sublevels, separated by \SI{\sim 13}{\mega\hertz}. The static quantization field of \SI{\sim5}{G} is oriented perpendicular to the trap surface.  Laser beams and RF control signals illuminate both ions uniformly.



The trapped-ion qubits are encoded such that all fundamental operations on the data qubit can be performed without disturbing the monitor qubit. The data qubit is encoded in the ground state ($g$-type) of the $^{40}$Ca$^+$ ion~ \cite{loschnauer2024scalable}. The monitor ion is encoded as a metastable qubit ($m$-type), and due to the relative $g$ factors of the ground and metastable states, it is $2.4$ times more sensitive to magnetic fields than the data qubit. The roles of monitor and data qubit are deterministically assigned to the first and second ion in the chain, respectively, using a spectrally-selective technique that eliminates the need for tightly-focused control fields~\cite{Nagerl1999, Piltz2014}, and is described here briefly: By applying appropriate voltages to the trapping electrodes, we orient the ion chain such that the monitor qubit incurs strong micromotion sidebands on the $S_{1/2} \leftrightarrow D_{5/2}$ transition, while the data ion remains on the RF null, where the micromotion sidebands are suppressed \cite{Shi2025, Lysne2024, Leibfried1999}.

A sequence of RF and 729-nm-laser pulses is used to prepare the monitor qubit in an equal superposition of the $m=+3/2$ and $m=-5/2$ states of the $D_{5/2}$ manifold; this is followed by RF pulses to prepare the data qubit in a superposition of the $m=\pm 1/2$ states of the $S_{1/2}$ manifold, as shown in \cref{fig:monitor-ion-qubits}. SK1 composite pulse sequences~\cite{Brown2004} are used to suppress over- and under-rotation errors on the micromotion transitions, resulting in a state preparation fidelity of approximately 99\%. The complete monitor/data qubit assignment procedure lasts approximately \SI{100}{\micro\second}. 



State detection of the qubits proceeds sequentially, as shown in \cref{fig:monitor-ion-qubits}. The data qubit is detected first by shelving one of the states to an empty level of the $D_{5/2}$ manifold (a state not used for the monitor qubit) and applying light resonant with the $S_{1/2} \leftrightarrow P_{1/2}$ transition, after which the data qubit is shelved to other empty $D_{5/2}$-manifold states. Following the monitor ion's second $\pi/2$ pulse (once again performed via a series of RF and 729-nm-laser pulses), the projection into the computational basis of its free evolution in the magnetic field is measured by de-shelving one of the states to the $S_{1/2}$ manifold and applying the readout light. The estimate of the magnetic field strength at realization $i$, $B_i$, is increased or decreased by a fixed amount inversely proportional to the Ramsey probe duration $\tau$ after each monitor-ion measurement, as $B_i = B_{i-1} \pm \alpha h / \tau \mu_\text{B}$, where $h$ is the Planck constant and $\mu_\text{B}$ is the Bohr magneton. The performance of the protocol is not strongly dependent on $\alpha$, and in these experiments we use the empirically-determined value of $\alpha = 0.05$. The drive field frequencies for the monitor and data qubits are updated (scaled by the appropriate Land\'{e} $g$-factors) after each realization to reflect the new estimate $B_i$. With a single monitor ion, each measurement is dominated by quantum projection noise \cite{Peik2006, Ludlow2015, Itano1993}, but performing many successive one-qubit measurements enables effective tracking of the magnetic field drift over time.


To quantify the effectiveness of the monitor qubit protocol over a range of noise strengths, we introduce  magnetic field fluctuations into the system. We apply noise with a power spectral density proportional to $1/f^2$, which is characteristic of slowly drifting parameters, such as ambient magnetic field fluctuations due to common environmental noise sources \cite{mills2022impact}. We study a range of five orders of magnitude in magnetic field noise power spectral density, from \SI{3.2}{(\micro G )^2 \per \hertz} to \SI{3.2e5}{(\micro G)^2 \per \hertz}. Note that the corresponding amplitude spectral density, which the magnetic field $B$ is proportional to, is described by $1/f$. In the rest of this paper we report amplitude spectral densities. The noise is applied by modulating the current in a pair of Helmholtz-configured coils inside the vacuum chamber that generate the quantization field, and is common to both ions. The conversion from coil current~$I$ to magnetic field~$B$, $\partial B / \partial I$, is calibrated using the ground-state Zeeman splitting of the $^{40}\text{Ca}^+$ ion. The large inductance of the coils acts as a low-pass filter with a \SI{2.4}{\hertz} cutoff frequency (extracted from auxiliary measurements). 


\begin{figure}
    \centering

    \includegraphics[width=1\linewidth]{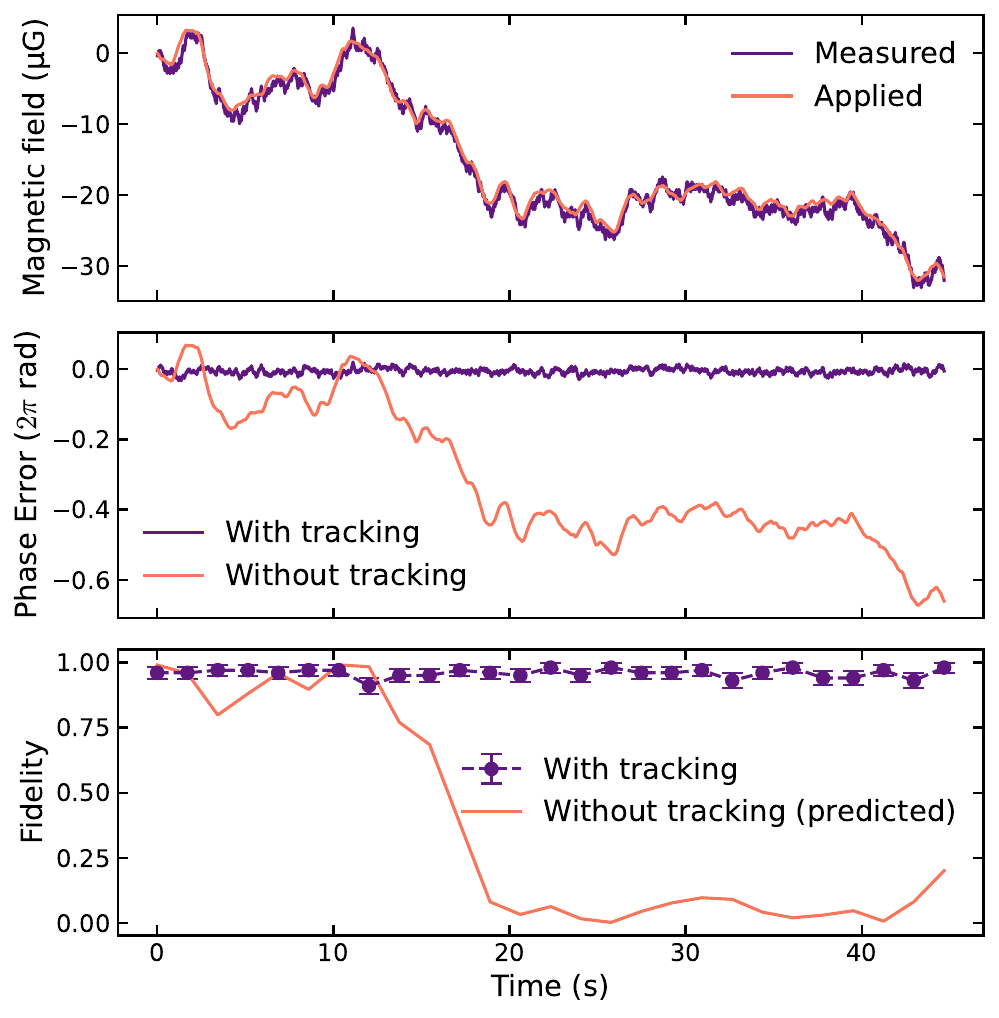}

    \caption{Tracking drifting magnetic fields with a monitor ion. \textbf{Top}: Magnetic field noise (applied via in-chamber coils in a Helmholtz configuration) calculated from applied current, as described in the text, and the measurements made by a monitor ion's magnetic field tracking protocol, plotted over \SI{45}{\second}. The applied noise strength is $\mathrm{ASD}_B(\SI{1}{\hertz}) = \SI{18}{\micro G \per \sqrt{\hertz}}$. \textbf{Center:} Calculated phase error accrued each realization by the data ion based on applied and measured magnetic field values. The orange trace is the estimated phase error the data ion would have accrued each realization if the monitor ion was not tracking the field, while the blue trace is the phase error with the monitor ion protocol active. These are calculated as described in the text. \textbf{Bottom:} Measured and estimated fidelities over the course of the run. Predicted fidelities are calculated as described in the text. The blue points are measured fidelities, calculated by binning the data-qubit measurements in groups of 100 realizations and taking the average probability of measuring the desired state. Error bars represent one standard deviation.}
    \label{fig:frequency-tracking}
\end{figure}



Through repeated, accurate measurement of the shared environment, the monitor ion protocol demonstrates the ability to maintain coherence in the presence of drifting magnetic fields. Standard recalibration methods using the data qubit alone would likely be infeasible for realistic noise strengths and experiment durations as used here, because the magnetic field quickly drifts far enough for the data qubit's phase to be entirely scrambled. The improvement in coherence is demonstrated over a \SI{45}{\second} time span in \cref{fig:frequency-tracking}. The top panel shows the close agreement of the applied magnetic field $\Delta B$ (taking into account the frequency response of the circuit) with the value extracted from the monitor ion protocol $B_i$. In the middle panel, the effects of this noise, in terms of phase errors accrued during a Ramsey experiment, are calculated. The orange curve shows the effect of the magnetic field drift on each realization of the data ion's experiment in the absence of drift tracking. This phase error $\delta \phi$ is calculated from the Bohr magneton $\mu_\mathrm{B}$, the applied magnetic field shift $\Delta B$, the data ion's interrogation time $\tau$, and the reduced Planck constant $\hbar$ as $\delta \phi =  2\mu_\mathrm{B} \Delta B \tau / \hbar$.  The phase error with tracking $\delta \phi_m$ is calculated similarly, but using the difference in magnetic field between the known applied field and the monitor ion's measurement, $\Delta B - B_i$. The bottom panel shows the effect of the protocol on experimental fidelities via Ramsey measurements on the data qubit. The orange curve shows the estimated fidelity if the monitor ion protocol was not active (calculated as $\cos^2(\delta \phi)$ for each realization) as well as the measured fidelities of the data qubit Ramsey sequences (using the monitor ion protocol) in the two-ion experiment.

These data show that at this relatively high noise strength (amplitude spectral density \SI{18}{\micro G \per \sqrt{\hertz}} at \SI{1}{\hertz}), the phase error without monitor ion correction approaches $\pi$ radians within less than a minute, leading to complete decoherence. In contrast, with the monitor qubit protocol activated, the additional accrued phase remains close to zero, and the coherence of the data qubit is maintained through 2700 experimental realizations over the course of the 45-second-long data collection run. To maintain high fidelity with this noise strength, recalibrations would have to happen so quickly as to approach the ``interleaved'' case where every realization of the desired algorithm would be followed by a realization of the recalibration measurements, as shown schematically in \cref{fig:monitor-ion-qubits}.  If the interleaved measurements happen at this cadence, and have durations similar to the data-qubit experiments, then the monitor ion protocol could double the throughput of the algorithm. Additionally, because the monitor-qubit measurements can occur simultaneously, rather than sequentially, with data-qubit experiments, the duty cycle of noise measurements is increased as well. When the data-qubit experiment and the noise measurements have similar durations, this leads to an expected approximately $\sqrt{2}$ improvement in the maximum experimental duration tolerable before the noise measurements can no longer track the drift. Alternatively, for a fixed probe time the protocol can tolerate a factor of approximately $\sqrt{2}$ higher amplitude spectral density (a factor of $2$ higher power spectral density) of magnetic-field noise before losing track of the drift. An additional benefit of using a monitor ion is that the effects of aliasing of higher-frequency noise are reduced with this decrease in dead time \cite{dick1989local, Ludlow2015}.

\begin{figure}
    \centering

    \includegraphics[width=\linewidth]{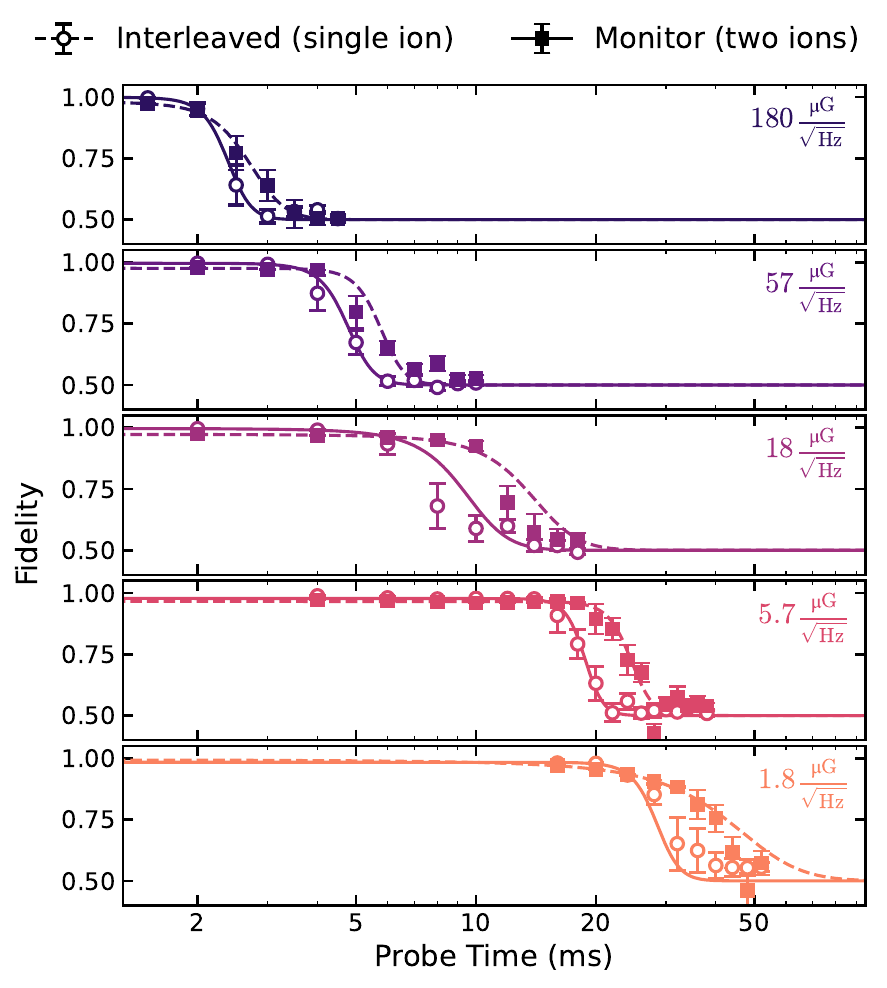}\\

    \caption{Averaged fidelity of data-qubit ($g$-qubit) Ramsey experiments as a function of probe time when using the monitor ion for real-time recalibration and when using interleaved field noise measurements with only a single ion. A range of magnetic field noise strengths are demonstrated, labeled in the upper right corner of each subplot. The monitor ion data show a maximum useful probe time of up to a factor of ${\sim}\sqrt{2}$ longer than the interleaved field-tracking experiments because of the higher noise-tracking duty cycle. Phenomenological fits to $\tanh$ (hyperbolic tangent) functions are used to extract the coherence times. Error bars represent one standard error on the mean computed from 10 one-minute realizations. The corresponding interleaved experiments use the same total number of realizations as the monitor-ion experiments.}
    \label{fig:probe-time-scan}
\end{figure}

\begin{figure}
    \centering
\includegraphics[width=0.9\linewidth]{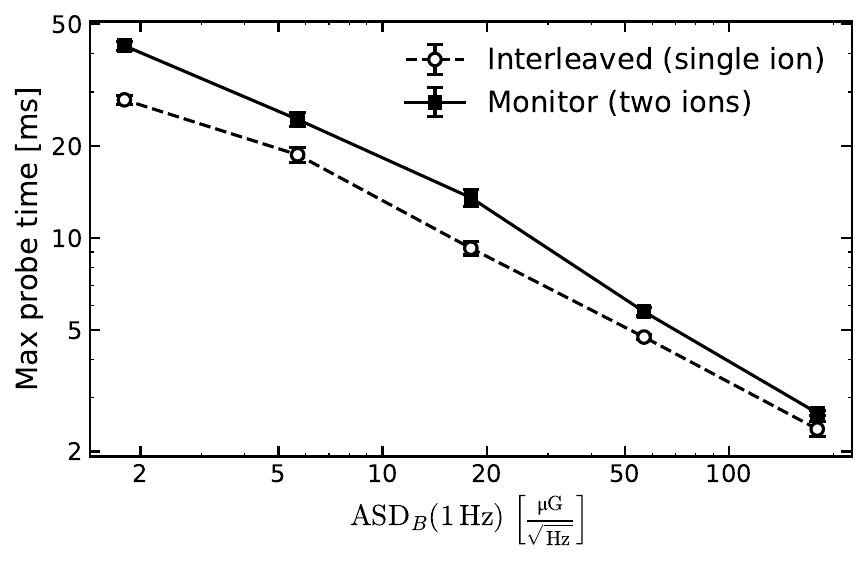}

    \caption{Maximum usable probe time as a function of the applied noise strength. The maximum probe time is the 50\% contrast (75\% fidelity) point of the fits to individual noise strength measurements (\cref{fig:probe-time-scan}). The monitor-ion protocol demonstrates consistently better performance, with the gains highest at low noise strengths due to the smaller fractional contribution of dead time. Error bars are statistical standard deviations from probe time fits (shown in \cref{fig:probe-time-scan}), and lines are guides to the eye.}
    \label{fig:noise-strength-scan}
\end{figure}

We further quantify the utility of monitor ions by measuring the maximum monitor Ramsey probe duration (and thus the maximum data-qubit experiment duration in our implementation) that can be used with high fidelity. At each probe time, approximately one minute worth of experimental realizations are performed as magnetic field noise is applied, with the applied magnetic field error starting at zero on the first realization and performing a random walk described by the noise spectrum, over that minute. This sequence is repeated 10 times, and the fidelity is taken to be the fraction of the realizations where the data qubit experiment resulted in a measurement of the intended state.  This can also be expressed in terms of a Ramsey fringe contrast $C$ as $(C+1)/2$. We also perform the comparable (interleaved) single-ion experiment, where only one ion is present and alternates between noise tracking and algorithm realizations as depicted in \cref{fig:monitor-ion-qubits}e. As is illustrated in \cref{fig:probe-time-scan}, the monitor ion probe times can be extended by approximately the expected factor of $\sqrt{2}$ over the interleaved probe times at the same noise strength. This improvement persists over a range of noise strengths but decreases for the strongest noise applied in these experiments (see~\cref{fig:noise-strength-scan}).

At strong noise strengths (and therefore short maximum probe durations), the benefits of the monitor ion decreases as the state preparation and measurement dead time becomes a more significant fraction of the total experiment duration. In these experiments, the monitor-ion Ramsey probe is longer than the data-qubit Ramsey probe by the duration of the state preparation and measurement pulses on the data qubit ($\approx\SI{1.1}{\milli\second}$). There is another dead time period of approximately the same duration for the monitor-ion state preparation and measurement, during which neither ion is being probed. For the single-ion interleaved measurements, the field-tracking probe time and the data-qubit experiment probe time durations are set equal to the corresponding times in the monitor-ion experiment for straightforward comparison.



\section{Discussion}

These results demonstrate the utility of monitor ions for addressing the problems of drifting parameters in quantum computers. While the addition of non-computational ions does represent a small overhead, many computing architectures already employ such extra ions for sympathetic cooling, entangled-pair generation, and mid-circuit state readout---with appropriate species choice and encoding, these existing additional ions can also serve as monitors. This, plus the crosstalk-free performance arising from dual-species or \textit{omg} architectures, means monitor qubits can be readily integrated into near-term quantum computers. The importance of such monitors will only increase as the depth of quantum circuits increases. Calibration and testing reportedly consumes up to 43\% of the total duty cycle of commercial trapped-ion quantum computing devices \cite{Maksymov2021}, and the number of parameters to calibrate only increases with larger system sizes.

The monitor ion protocol implementation demonstrated here is limited to tracking of noise characterized by a drift slow compared to the timescale of a single experimental realization. This limit on noise speed is due to the similar sensitivities of the monitor ion and data ion to the magnetic field.  Tracking of faster noise would be possible with a modified system, where the monitor ions are able to acquire information at a rate much faster than the data ion. This increase can be achieved either by reducing the quantum projection noise with a larger number of monitor ions \cite{Singh2023}, or by using a monitor ion with a greater magnetic field sensitivity than that of the data ion, such as in a species with first-order magnetic-field-insensitive transitions. In these cases, as shown by \cite{Majumder2020}, it is theoretically possible to keep data-qubit error below a fixed threshold, by means of feedforward from a monitor qubit, for arbitrarily long times.

These protocols can also be extended beyond measurements of global magnetic field fluctuations to other parameters of interest. Monitor ions can also potentially be used to track laser intensity fluctuations \cite{Majumder2020}, laser phase fluctuations \cite{Zheng2024, Kim2023, saner_breaking_2023}, magnetic field gradients \cite{Gupta2020a}, or other drifting parameters. Of particular interest for trapped-ion systems would be monitor ions that locally track trap-frequency fluctuations induced by the charging of dielectrics from lasers \cite{Wang2011, Jung2023}; these frequency fluctuations are an increasingly important limitation as ion traps with integrated control technologies become more prevalent \cite{Lee2024, Mehta2020, Ivory2021}.

\section*{Acknowledgments}

This material is based upon work supported by the National Science Foundation Graduate Research Fellowship under Grant No. 2141064. This material is based upon work supported under Air Force Contract No. FA8702-15-D-0001 or FA8702-25-D-B002. Any opinions, findings, conclusions or recommendations expressed in this material are those of the author(s) and do not necessarily reflect the views of the U.S. Air Force.

\bibliographystyle{apsrev4-2}
\bibliography{refs, refs_exported_from_ads}

\end{document}